\documentclass[preprint,prd,showpacs]{revtex4} 
 
\begin{document} 
\input{epsf} 
   
%\begin{flushright}    
%\vspace*{-2cm}  
%gr-qc/yymmnnn \\  
%April 1, 2007 \\     
%\vspace*{1cm}    
%\end{flushright}    

\title{Energy Density-Flux Correlations \\
in an Unusual Quantum State and in the Vacuum}

\author{ L.H. Ford}
 \email[Email: ]{ford@cosmos.phy.tufts.edu} 
 \affiliation{Institute of Cosmology  \\
Department of Physics and Astronomy\\ 
         Tufts University, Medford, MA 02155}
\author{Thomas A. Roman}
  \email[Email: ]{roman@ccsu.edu}
  \affiliation{Department of Mathematical Sciences \\
 Central Connecticut State University \\  
New Britain, CT 06050} 
\date{\today}

\begin{abstract} 
In this paper we consider the question of the degree to which  
negative and positive energy are intertwined. We examine in more 
detail a previously studied quantum state 
of the massless minimally coupled scalar field, which we call a ``Helfer state''. 
This is a state in which the energy density can be made arbitrarily negative 
over an arbitrarily large region of space, but only at one instant in time. 
In the Helfer state, the negative energy density is accompanied by rapidly 
time-varying energy fluxes. It is the latter feature which allows the quantum 
inequalities, bounds which restrict the magnitude and duration of negative energy, 
to hold for this class of states. An observer who initially passes through the 
negative energy region will quickly encounter fluxes of positive energy which 
subsequently enter the region. We examine in detail the correlation between the energy 
density and flux in the Helfer state in terms of their expectation values. We then 
study the correlation function between energy density and flux in the Minkowski vacuum 
state, for a massless minimally coupled scalar field in both two and four dimensions. 
In this latter analysis we examine correlation 
functions rather than expectation values. Remarkably, we see qualitatively similar behavior 
to that in the Helfer state. More specifically, an initial negative energy vacuum fluctuation 
in some region of space is correlated with a subsequent flux fluctuation of positive energy into 
the region. We speculate that the mechanism which ensures that the quantum inequalities 
hold in the Helfer state, as well as in other quantum states associated with negative energy, 
is, at least in some sense, already ``encoded'' in the fluctuations of 
the vacuum.

\end{abstract}


\maketitle 
 
\baselineskip=14pt 

\section{Introduction}
It has been known for quite some time that quantum field theories 
generically contain states associated with negative energy \cite{EGJ}. 
Indeed, negative energy seems to be required by the laws of physics for such 
effects as Hawking evaporation (which allows the unification of black holes 
with the laws of thermodynamics) \cite{H75}, and for the stability of the Minkowski vacuum. 
In the former case, the positive Hawking radiation observed at infinity is paid 
for by a flux of negative energy down the horizon of the black hole \cite{DFUC}. In the latter case, 
since the vacuum is not an eigenstate of energy density, there will be energy 
density fluctuations in the vacuum. In order for the averaged energy density to be zero, 
there must exist negative, as well as positive, energy density vacuum fluctuations. States 
associated with negative energy are now routinely produced in the laboratory, e.g, squeezed 
vacuum states and the Casimir effect \cite{SY,C}. However, the energy density itself 
is too small to be directly measurable.

On the other hand, unrestricted amounts of negative energy could result in 
gross macroscopic effects such as violation of the second law of 
thermodynamics \cite{F78,F91}, and exotic spacetime geometries 
such as wormholes \cite{MT,MTY}, warp drives \cite{A,PFWD,ER,K98}, and time machines \cite{H92}. 
However, the laws of quantum field theory contain restrictions on negative energy in the 
form of ``quantum inequalities''. These involve bounds on the magnitude of negative energy and 
on its temporal or spatial extent. There has been much progress 
in this area in the last 15 years (for some recent reviews see 
Refs.~\cite{TR-MGM10,CJF-Rev,LF100}). 

Some years ago, Helfer~\cite{Helfer96} suggested the existence of a class of quantum states in 
which one could make the energy density arbitrarily negative in an arbitrarily 
large region of space, even in Minkowski spacetime. We call these ``Helfer states.'' 
The present authors, together with Helfer, verified this claim in an earlier paper 
\cite{FHR}, which we will refer to as FHR. 
In that paper the characteristics of the energy density were 
analyzed. We argued that a crucial feature of these states is the fact 
that although the energy density can be made arbitrarily negative over an arbitrarily 
large region of space, this can be done {\it only at one instant of time}. The 
worldline quantum inequalities must hold for these states as well, since they have 
been proven to hold for {\it all} quantum states. Therefore, for this to be true, there 
must be rapidly time-varying fluxes which accompany the negative energy density. 
In the current paper, we show that 
this is indeed the case. The expectation values of the energy density and flux 
in a particular Helfer state are calculated and graphed, and their correlations are studied.  
An inertial observer who initially passes through the negative energy region must 
quickly encounter fluxes of positive energy which subsequently enter the region, thus 
ensuring that the time-averaged sampled energy density along the observer's worldline is 
non-negative, in accordance with the quantum inequalities. 

These fascinating correlations of flux and energy density are by no means unique to 
the Helfer states. In the second part of the paper we calculate the energy density-flux 
correlation function for the Minkowski {\it vacuum} state, in both two and four-dimensional spacetime. 
Although here we analyze correlation functions, as opposed to expectation values, nonetheless, 
we find remarkably similar behavior for the fluctuations of energy density and flux 
in the vacuum to their expectation values in the Helfer states. An initial negative energy 
vacuum fluctuation in some region of space is correlated with a subsequent flux fluctuation 
of positive energy into the region. We speculate that the mechanism which ensures 
that the quantum inequalities hold in the Helfer state, as well as in other quantum states 
associated with negative energy, is, at least in some sense, already ``encoded'' in the 
fluctuations of the vacuum.

\section{The Helfer State} 
\label{sec:char} 

Here we briefly summarize the results of FHR, and refer the reader to
that paper for more details. The quantum state used in FHR is a
superposition of two-particle states, which can be expressed as
\begin{equation} 
|\psi \rangle = N \left[ |0\rangle + \int d^3k_1\, d^3k_2\, b({\bf k}_1,{\bf k}_2)
\, |{{\bf k}_1,{\bf k}_2} \rangle \right] \,, 
                                           \label{eq:state} 
\end{equation} 
where the two-particle state, $|{{\bf k}_1,{\bf k}_2}\rangle $,
contains particles with three-momenta ${\bf k}_1$ and ${\bf k}_2$. The
normalization factor $N$ is given by
\begin{equation} 
N = \left[1 + 2 \int d^3k_1\, d^3k_2\, |b({\bf k}_1,{\bf k}_2)|^2  
              \right]^{-\frac{1}{2}} \,.       \label{eq:norm} 
\end{equation} 
We consider the case where
\begin{equation} 
b({\bf k}_1,{\bf k}_2) = \chi({\bf k}_1+{\bf k}_2) \,  
   (|{\bf k}_1||{\bf k}_2|)^{-1}\, ,    \label{eq:b} 
\end{equation} 
and
\begin{equation}  
\chi({\bf p}) = \left\{ \begin{array}{ll}  
                        \chi_0 \,, & \mbox{if $|{\bf p}| \leq p_0$}  \\ 
                         0\,,     & \mbox{otherwise}         
                         \end{array}  \right.  
\label{eq:chi}    
\end{equation} 
where $\chi_0$ and $p_0$ are arbitrary constants, and 
${\bf p} = {\bf k}_1+{\bf k}_2$. (This is the $\nu =
-1/2$ limit of the state discussed in FHR.) In addition, we require
that the magnitudes of the momenta of the particles be less than 
a cutoff parameter,
$\Lambda$, so that $|{\bf k}_1|, |{\bf k}_2| < \Lambda$. Thus, the
state is described by the three parameters, $\Lambda$, $\chi_0$, 
and $p_0$. In the limit that $\Lambda$ becomes large, with $\chi_0$ 
and $p_0$ fixed, the two members of the pair of particles have nearly
opposite momenta. This is the limit of particular interest. 

\subsection{Energy Density in the Helfer State}

The energy density of a massless, minimally coupled scalar field may
be obtained from the renormalized two-point function using 
\begin{equation} 
\rho = \frac{1}{2}  
\lim_{{{\bf x'}\rightarrow {\bf x}}\atop {t'\rightarrow t}}\left[ 
(\partial_t \partial_{t'} + \mbox{\boldmath $\nabla \cdot \nabla'$})\, 
 \langle :\varphi(x)\, \varphi(x'):\rangle \right] \,. 
\end{equation}
Units in which $\hbar=c=1$ are used throughout this paper. 
In our case, this two-point function may be shown to be
\begin{eqnarray}  
\langle :\varphi(x)\, \varphi(x'):\rangle &=& \frac{2 N^2}{(2 \pi)^3} \, 
{\rm Re} \int  d^3k\, d^3k'\, \frac{1}{\sqrt{\omega \omega'}}\, 
                                                       \nonumber \\ 
&\times&\Bigl[ 2 {\rm e}^{i({\bf k'}\cdot{\bf x'}-{\bf k}\cdot{\bf x})} 
{\rm e}^{i(\omega\, t - \omega'\, t')} \,  
\int d^3k_1 b^*({\bf k}_1,{\bf k}) b({\bf k}_1,{\bf k'}) \nonumber \\ 
 &+&  {\rm e}^{i({\bf k}\cdot{\bf x}+{\bf k'}\cdot{\bf x'})} 
{\rm e}^{-i(\omega\, t + \omega'\, t')}\, b({\bf k},{\bf k'}) \Bigr] \,. 
                                                      \label{eq:2pt} 
\end{eqnarray} 
The resulting energy density is
\begin{eqnarray} 
\rho &=& \frac{ N^2}{(2 \pi)^3} \, {\rm Re} \int  d^3k\, d^3k'\, 
\sqrt{\omega \omega'} \, (1+ \hat{{\bf k}}\cdot\hat{{\bf k}}')\, 
\Bigl[ 2 {\rm e}^{i({\bf k'}-{\bf k})\cdot{\bf x}} 
{\rm e}^{i(\omega  - \omega')t} \,  
\int d^3k_1 b^*({\bf k}_1,{\bf k}) b({\bf k}_1,{\bf k'})  
                                  \nonumber \\     
&&\,\,\,\,- {\rm e}^{i({\bf k}+{\bf k'})\cdot{\bf x}} 
{\rm e}^{-i(\omega + 
\omega') t} \, b({\bf k},{\bf k'}) \Bigr] \,,   \label{eq:rho} 
\end{eqnarray}  
where $\hat{{\bf k}}$ and $\hat{{\bf k}}'$ are unit vectors in the directions 
of ${\bf k}$ and ${\bf k'}$, respectively. It is convenient to write
$\rho = \rho_1 + \rho_2$, where $\rho_1$ is the contribution quadratic
in $b$, and $\rho_2$ is that linear in $b$.  

In general, it is difficult to evaluate $\rho$ explicitly for the
choice of $b({\bf k}_1,{\bf k}_2)$ given by Eqs.~(\ref{eq:b}) and
 (\ref{eq:chi}). However, in FHR approximate forms were derived in the 
limit that the dominant contribution comes from modes whose frequency
is large compared to $p_0$, which is expected to be the case when
$\Lambda \gg p_0$. These approximate forms contain an undetermined
constant $q \gg 1$ which is introduced by making a lower cutoff in the
$\omega$-integrals of $q p_0$.
The result for $\rho_2$ is
\begin{equation} 
\rho_2 \approx -\frac{\chi_0 N^2}{6 \pi^2} \,f_2(p_0,r)\,  
 \int_{q p_0}^{\Lambda} d\omega \, \omega^{-1}\, \cos (2 \omega t) \,, 
                                                   \label{eq:rho2b} 
\end{equation}
where
\begin{equation} 
f_2(p_0,r) = \frac{4\pi}{r^5}\,  
[3(p_0^2 r^2-2)\sin (p_0 r) - p_0 r(p_0^2 r^2-6)\cos (p_0 r)] 
  \,.                                        \label{eq:f} 
\end{equation} 
The function $f_2(p_0,r)$ has the following behavior: 
\begin{equation} 
f_2(p_0,r) \sim \left\{ \begin{array}{ll}  
                        -(4 \pi /r^2) \,{p_0}^3 \, {\rm cos} \, p_0 r \,,  
                      & \mbox{$r \gg {p_0}^{-1}$}  \\ 
                         (4\pi/5) \, p_0^5 \,, & \mbox{$ r \ll {p_0}^{-1}$}         
                         \end{array}  \right. \,. 
\label{eq:f2sim} 
\end{equation} 
At $t=0$, Eq.~(\ref{eq:rho2b}) becomes 
\begin{equation} 
\rho_2 \approx -\frac{\chi_0 N^2}{6 \pi^2} \,f_2(p_0,r)\,  
\ln \left( 
\frac{\Lambda}{q p_0} \right) \,. 
\label{eq:rho2c} 
\end{equation} 

Let us now examine the behavior of $\rho_1$. From Eq. (33) of FHR, 
we have that 
\begin{eqnarray} 
\rho_1  &=& \frac{2\,{\chi_0}^2 \, N^2}{(2 \pi)^3} \,  
{\rm Re} \int d^3k_1\, {|k_1|}^{-2} \,  
\int_{|{\bf p}|\leq p_0} d^3p\, \int_{|{\bf p}'|\leq p_0} d^3p'\,  
(1+ \hat{{\bf k}}\cdot\hat{{\bf k}}')\, 
\nonumber \\ 
\,\,\,\,&&\times {|{\bf p}-{\bf k}_1|}^{-1/2} \,{|{\bf p}'-{\bf k}_1|}^{-1/2} \, 
{\rm e}^{-i({\bf p}-{\bf p}')\cdot{\bf x}}  
\, {\rm e}^{i[(|{\bf p}-{\bf k}_1| - |{\bf p}'-{\bf k}_1|)t]}\,, 
\label{eq:rho_1_2} 
\end{eqnarray} 
where ${\bf p} = {\bf k}_1 +{\bf k}$ and ${\bf p}' = {\bf k}_1 +{\bf k}'$.
In the high frequency limit, $\rho_1$ at $t=0$ is given by
\begin{equation} 
\rho_1 \approx \frac{2 \, {\chi_0}^2 N^2}{\pi^2} \,f_1(p_0,r)\,  
\ln \left( 
\frac{\Lambda}{q p_0} \right) \,, 
\end{equation} 
where
\begin{equation} 
f_1(p_0,r) \equiv \frac{{(4\pi)}^2}{r^6}\,  
{[\sin (p_0 r) - p_0 r \,\cos (p_0 r)]}^2 \, , 
\label{eq:f1} 
\end{equation} 
and  
\begin{equation} 
f_1(p_0,r) \sim \left\{ \begin{array}{ll}  
                        {(4 \pi p_0)}^2 /r^4 \, {\rm cos}^2 \, (p_0 r) \,,  
                      & \mbox{$r \gg {p_0}^{-1}$}  \\ 
                         {(4\pi /3)}^2 \, p_0^6 \,, & \mbox{$ r \ll {p_0}^{-1}$}   
                         \end{array}  \right. \,. 
\label{eq:f1sim} 
\end{equation} 
The time dependence of $\rho_1$ is given by the factor of
$ {\rm e}^{i[(|{\bf p}-{\bf k}_1| - |{\bf p}'-{\bf k}_1|)t]}$ in
Eq.~(\ref{eq:rho_1_2}). In the high frequency limit,
\begin{equation}
|{\bf p}-{\bf k}_1| = k_1\, \sqrt{1 -
2 \frac{{\bf p}\cdot {\bf \hat{k}}_1} { k_1} + \frac{p^2}{k_1^2} }
\approx k_1 - {\bf p}\cdot {\bf \hat{k}}_1 \, ,  
\end{equation}
where $k_1 = |{\bf k}_1|$. As a result,
\begin{equation}
\omega -\omega' = |{\bf p}-{\bf k}_1|- |{\bf p}'-{\bf k}_1|
\approx {\bf \hat{k}}_1 \cdot ({\bf p}'-{\bf p}) \,. \label{eq:wwp}
\end{equation}
Because $|{\bf p}|, |{\bf p}'| \leq p_0$, the time dependence of
$\rho_1$
is on a time scale of order $1/p_0$, and can be neglected if we are 
interested only in times for which $|t| \ll 1/p_0$. The time variation
of the regions of negative energy is governed by the rapid time
dependence of $\rho_2$ coming from the $ \cos (2 \omega t)$ factor
in Eq.~(\ref{eq:rho2b}). This is the time variation which allows the
quantum inequalities to be satisfied in this case.

The integral in the normalization factor,  Eq.~(\ref{eq:norm}), is
\begin{equation} 
I =\chi_0^2 \,\int d^3k\,\int_{|{\bf p}| \leq p_0} d^3p\,  
(|{\bf k}||{\bf k - p}|)^{-2} \,, 
\end{equation} 
where ${\bf k} = {\bf k}_1$ and ${\bf p} = {\bf k}_1+{\bf k}_2$.
In the limit where $|{\bf k}_1| \gg p_0$, 
$(|{\bf k}||{\bf k - p}|)^{-2} \approx |{\bf k}|^{-4}$, so we can
write
\begin{equation} 
I \approx \chi_0^2 \,\int d^3k \, |{\bf k}|^{-4}\, \int_{|{\bf p}| \leq p_0} d^3p\,  
  =  4 \pi \chi_0^2  \int_{q p_0}^\Lambda \, \omega^{-2} \, d\omega 
\,\int_{|{\bf p}| \leq p_0} d^3p =  
\frac{16 \pi^2}{3} \chi_0^2 \,
\biggl(\frac{{p_0}^2}{q}-\frac{{p_0}^3}{\Lambda}\biggr)\,. 
\end{equation} 
When $\Lambda \gg q p_0$, this becomes
\begin{equation} 
I \approx  \frac{16 \pi^2}{3} \chi_0^2 \,
\frac{{p_0}^2}{q} \,,
\end{equation}
which leads to 
\begin{equation}
N \approx {\biggr(1+\frac{32 \pi^2}{3} \chi_0^2 \,
\frac{{p_0}^2}{q}\biggr)}^{-1/2} \,.
\end{equation}
 
For any value of $p_0$, the energy density is  
approximately constant in space over a region of size, $r \alt {p_0}^{-1}$.  
In this region, $(f_2/f_1) \approx  
9/(20 \, \pi \, p_0)$. If we choose  
\begin{equation} 
\chi_0 <  \frac{3}{80 \, \pi \, p_0} \,, \label{eq:chi_lim}
\end{equation} 
then $|\rho_2| / \rho_1 > 1$  and hence $\rho < 0$ in this region. In the limit that
$\Lambda \rightarrow \infty$, we can make $\rho$ arbitrarily negative
in this region. Furthermore, by making $p_0$ small, we can make this
region arbitrarily large. Note that if Eq.~(\ref{eq:chi_lim}) is
satisfied, and if $q \gg 1$, then $N \approx 1$. In this case, the
unknown parameter $q$ and the cutoff parameter $\Lambda$ only appear 
in the energy density in the ratio $\Lambda/q$. Thus we can now take
this ratio to be the effective cutoff. 

\begin{figure} 
\begin{center}
\leavevmode\epsfysize=8cm\epsffile{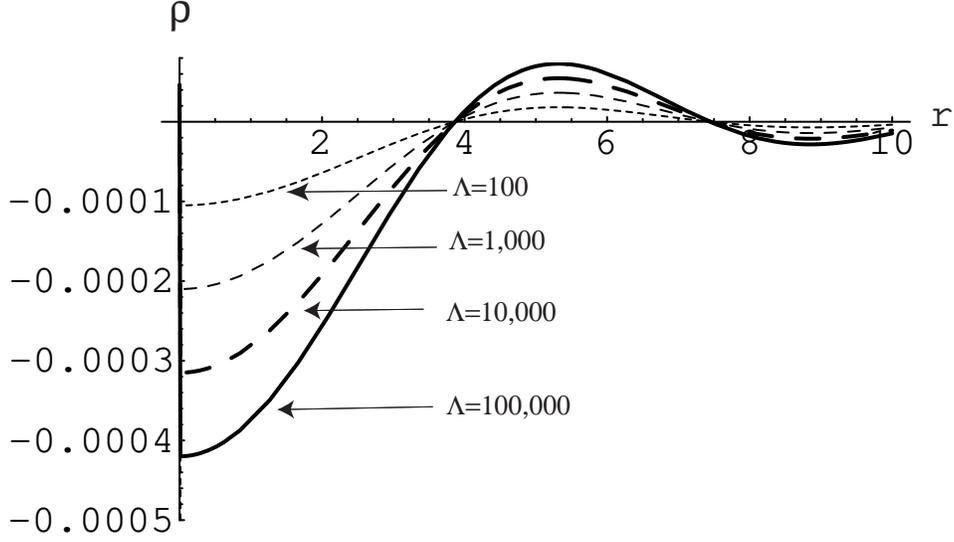} 
\end{center} 
\caption{The total energy density $\rho=\rho_1+\rho_2$, 
is plotted as a function of $r$, at $t=0$, for varying $\Lambda$. 
We have chosen $p_0=1$ and $q=10$. The energy density becomes 
more negative with increasing $\Lambda$. Note that on scales where 
$r \alt {p_0}^{-1}=1$, the energy density is approximately constant 
and negative. Such a region may be made as large as one likes by 
decreasing $p_0$.} 
\label{fig:ED-Fig1} 
\end{figure}

\begin{figure} 
\begin{center} 
\leavevmode\epsfysize=8cm\epsffile{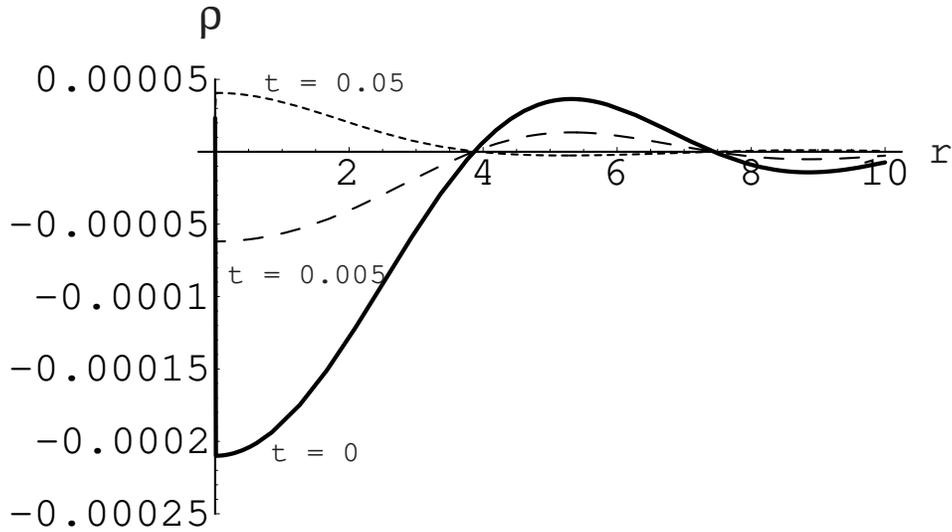}
\end{center} 
\caption{The total energy density $\rho=\rho_1+\rho_2$, 
as a function of $r$, for varying $t$. 
We have chosen $\Lambda=1000$, $p_0=1$ and $q=10$. At $t=0$ the energy 
density in the region near $r=0$ is negative. 
By $t=0.005$, it has become substantially 
less negative, and by $t=0.05$ it has already become positive.} 
\label{fig:ED-Fig2} 
\end{figure}

For the purpose of plotting the energy density, we let $q = 10$ and
set
\begin{equation} 
\chi_0 = \frac{3}{800 \, \pi \, p_0} \,,
\end{equation}  
which ensures that there is a region of negative energy about $r=0$.
In Fig.~\ref{fig:ED-Fig1}, we show the energy density as a function of
$r$ at $t=0$ for several values of $\Lambda$. In
Fig.~\ref{fig:ED-Fig2}, we illustrate the energy density for one
choice of $\Lambda$ at several values of $t$ as functions of $r$. The
special feature of this quantum state is that it exhibits large
negative energy density at $t=0$, but this negative energy rapidly
disappears, as required by the quantum inequalities. The time scale
for this change is small compared to the light travel time across the
negative energy region. This will be discussed in more detail
in the next subsection, where we show that 
this rapid change in $\rho$ is associated with large energy fluxes.

\subsection{The Energy Flux in the Helfer State}

Let us temporarily switch to Cartesian coordinates. Then in the $z$-direction 
we have
\begin{equation}
\langle T_{0z} \rangle = - \langle T^{0z} \rangle 
=  \frac{1}{2} \, 
\lim_{{{\bf x'}\rightarrow {\bf x}}\atop {t'\rightarrow t}}\left[ 
(\partial_t \partial_{z'} + \partial_{t'} \partial_z)\, 
 \langle :\varphi(x)\, \varphi(x'):\rangle \right] \,.
\end{equation}
Using Eq.~(\ref{eq:2pt}), we obtain 
\begin{equation}
\langle T_{0z} \rangle = I_1 +I_2 \, ,   \label{eq:T_0z} 
\end{equation} 
where 
\begin{equation}
I_2 = \frac{ N^2}{(2 \pi)^3} \, {\rm Re} \int  d^3k\,\int d^3k'_1 \,
\frac{(\omega {k'}_z + \omega' {k}_z)}{\sqrt{\omega \omega'}} \,
{\rm e}^{i({\bf k'}+{\bf k})\cdot{\bf x}} 
{\rm e}^{-i(\omega  + \omega')t}\, b({\bf k},{\bf k'})\, ,
\label{eq:I_2}
\end{equation}
and 
\begin{eqnarray}
I_1 &=& - \frac{2 N^2}{(2 \pi)^3} \, {\rm Re} \int  d^3k\,\int d^3k' \,
\frac{(\omega {k'}_z + \omega' {k}_z)}{\sqrt{\omega \omega'}} \,
{\rm e}^{i({\bf k'}-{\bf k})\cdot{\bf x}} 
{\rm e}^{i(\omega  - \omega')t} \, \nonumber \\ 
&\times& \, \int d^3k_1 b({\bf k}_1,{\bf k}) b({\bf k}_1,{\bf k'})\,.
\label{eq:I1}
\end{eqnarray}
In the last expression, we use the fact that  $b({\bf k},{\bf k'})$ 
is real. At $t=0$, we have
\begin{eqnarray}
I_1(t=0) &=& - \frac{2 N^2}{(2 \pi)^3} \,  \int  d^3k\,\int d^3k' \,
\frac{(\omega {k'}_z + \omega' {k}_z)}{\sqrt{\omega \omega'}} \,
\cos[({\bf k'}-{\bf k})\cdot{\bf x}]  \, \nonumber \\ 
 &\times& \, \int d^3k_1 b({\bf k}_1,{\bf k}) b({\bf k}_1,{\bf k'})\,.
\label{eq:I10}
\end{eqnarray}
However, $I_1(t=0) =0$, as may be seen by making the replacements
${\bf k} \rightarrow -{\bf k}$, ${\bf k}' \rightarrow -{\bf k}'$, and
${\bf k}_1 \rightarrow -{\bf k}_1$ and using the fact that 
$b({\bf k}_1,{\bf k}) = b(-{\bf k}_1,-{\bf k})$. We can see from 
Eq.~(\ref{eq:wwp}) that $I_1$ will only vary on a time scale of order
$1/p_0$, as does $\rho_1$. Thus, if we are interested in times close
to $t=0$, we may set $I_1 \approx 0$.

Next we proceed as in FHR and examine $I_2$. Using ${\bf p} = {\bf k}+{\bf k'}$, 
and taking the limit in which $\omega$ becomes large compared to
$p_0$, we 
can expand 
the prefactor in $I_2$ as follows: 
\begin{eqnarray} 
\frac{(\omega {k'}_z + \omega' {k}_z)}{\sqrt{\omega \omega'}} &=&  
\sqrt{|{\bf k}| \,|{\bf p}-{\bf k}|} \, 
\left [\frac{({\bf p}-{\bf k})\cdot \hat{\bf z}} 
{\sqrt{({\bf p}-{\bf k})\cdot({\bf p}-{\bf k})}} + 
\hat{\bf k}\cdot \hat{\bf z} \right ] \nonumber  \\
&=&\sqrt{|{\bf k}| \,|{\bf p}-{\bf k}|}\,
\left [\left(\frac{p_z}{\omega} - \hat{\bf k}\cdot \hat{\bf z}\right) 
\left(1 -\frac{2 \,\hat{{\bf k}}\cdot{\bf p}}{\omega} +\frac{p^2}{\omega^2}  
                                                          \right)^{-\frac{1}{2}} 
+\hat{\bf k}\cdot \hat{\bf z} \right]
\nonumber  \\  
&\approx& p_z - (\hat{\bf k}\cdot {\bf p}) \hat{k}_z + \cdots \,, 
\end{eqnarray} 
where $\hat{k}_z \equiv \hat{\bf k}\cdot \hat{\bf z}$. So we can now write 
the integral $I_2$ as
\begin{equation}
I_2 \approx \frac{\chi_0 N^2}{(2 \pi)^3} \, {\rm Re}\int d^3k\, 
\int_{|{\bf p}|\leq p_0} d^3p\, 
[p_z - (\hat{\bf k}\cdot {\bf p}) \hat{k}_z] 
\, {\rm e}^{i{\bf p}\cdot{\bf x}}\, \, \omega^{-2} \,{\rm e}^{-2 i \omega t} \,. 
\end{equation}
Now let 
\begin{equation}
I_2 = I_{2a}+I_{2b} \,,
\end{equation}
where
\begin{equation}
I_{2a} \equiv \frac{\chi_0 N^2}{(2 \pi)^3} \, {\rm Re}\int d^3k\, 
\int_{|{\bf p}|\leq p_0} d^3p\, \,
p_z \, {\rm e}^{i{\bf p}\cdot{\bf x}}\, \, \omega^{-2} \,{\rm e}^{-2 i \omega t} 
\end{equation}
and
\begin{equation}
I_{2b}  \equiv -\frac{\chi_0 N^2}{(2 \pi)^3} \, {\rm Re}\int d^3k\, 
\int_{|{\bf p}|\leq p_0} d^3p\,\, 
(\hat{\bf k}\cdot {\bf p}) \hat{k}_z 
\, {\rm e}^{i{\bf p}\cdot{\bf x}}\, \, \omega^{-2} \,{\rm e}^{-2 i \omega t} \,. 
\end{equation}

First let us evaluate $I_{2a}$. Performing the angular integrations, we have
\begin{equation} 
\int d^3k = 4 \pi \int d\omega \, \omega^2 
\end{equation} 
and
\begin{equation}
I'_{2a} \equiv \int_{|{\bf p}|\leq p_0} d^3p\, \,
p_z \, {\rm e}^{i{\bf p}\cdot{\bf x}}
=\int_{|{\bf p}|\leq p_0} d^3p\, \,
p_z \,{\rm e}^{i p r {\rm cos}\theta} \,,
\label{eq:I'2a}
\end{equation} 
where $r=|{\bf x}|$ and $\theta$ is the angle 
between $\bf x$ and $\bf p$. 

We are assuming spherical symmetry, so let us choose, 
without loss of generality, $\bf x$ to point along the 
$z$-axis. Then $p_z = {\bf p}\cdot {\hat {\bf z}} = p\, {\rm cos}\theta$, 
where $p = |{\bf p}|$. We can now write 
\begin{equation}
I'_{2a} = \int_{|{\bf p}|\leq p_0} d^3p\, \,
p_z \,{\rm e}^{i p r {\rm cos}\theta} \
= 2 \pi \int_0^{p_0}  dp \, p^3 \int_{-1}^1 dc 
\,c \, {\rm e}^{i p r c} 
= - i \, g_2(p_0, r) \, ,
\label{eq:I'2a_2nd}
\end{equation}  
where $c = \cos\theta$ and  
\begin{equation} 
g_2(p_0, r) = \frac{4 \pi }{r^4}\,
[3 p_0 r \,{\rm cos}(p_0 r)+({p_0}^2 r^2-3) \, {\rm sin}(p_0r)] \,.
\label{eq:g_2}
\end{equation} 
Therefore $I_{2a}$ becomes
\begin{equation}
I_{2a} = -\frac{\chi_0 N^2}{(2 \pi)^2} \, g_2(p_0, r) 
\int_{q p_0}^{\Lambda} d\omega \, \sin (2 \omega t) \,,   
\end{equation} 
where as before, $q$ is a constant chosen so that the approximations made in finding  
the integrand ($\omega \gg p_0$) are valid throughout the range of integration. 
Evaluating the integral we have
\begin{equation}
I_{2a} = \frac{\chi_0 N^2}{4 \pi^2} \, g_2(p_0, r) 
\left[\frac{{\rm cos}(2 \Lambda t)- {\rm cos}(2 q p_0 t)}{t}\right] \,.
\label{eq:I_2a_final}
\end{equation} 

We now turn our attention to $I_{2b}$. First fix the direction 
of ${\bf p}$ and integrate over the directions 
of ${\bf k}$.  In a right-handed $x,y,z$ coordinate system, 
define: $\theta'$ to be the angle between $\bf k$ and the $z$-axis, 
$\theta$ to be the angle between $\bf p$ and the $z$-axis, 
$\phi'$ to be the azimuthal angle in the $x,y$ plane for $\bf k$, 
$\phi$ to be the azimuthal angle in the $x,y$ plane for $\bf p$, and 
$\gamma$ to be the angle between $\bf k$ and $\bf p$. Then we have that 
${ \hat k}_z ={\bf \hat k} \cdot {\bf \hat z} = {\rm cos}\theta', 
{\bf \hat k} \cdot {\bf p} = |{\bf p}| \, {\rm cos}\gamma$, and 
${\bf p} \cdot {\bf x}= p\, r\, {\rm cos}\theta$. Making use of the
fact that~\cite{Jackson}
\begin{equation}
{\rm cos}\gamma = {\rm cos}\theta \, {\rm cos}\theta' + 
{\rm sin}\theta \,{\rm sin}\theta' \,{\rm cos}(\varphi-\varphi') \,,
\end{equation}  
we can write  
\begin{eqnarray}
I_{2b} & = & -\frac{\chi_0 N^2}{(2 \pi)^3} \, {\rm Re}
\int_{|{\bf p}|\leq p_0} d^3p\,\, {\rm e}^{i{\bf p}\cdot{\bf x}}\,
\int d^3k\,p \,{\rm cos}\gamma \, {\rm cos}\theta' 
\,\, \omega^{-2} \,{\rm e}^{-2 i \omega t} \nonumber\\
&=&  -\frac{\chi_0 N^2}{(2 \pi)^3} \, {\rm Re}
\int_{|{\bf p}|\leq p_0} d^3p\,\, {\rm e}^{i{\bf p}\cdot{\bf x}}\,
\frac{4 \pi}{3} p \,{\rm cos}\theta \, \int \, d\omega
 \,{\rm e}^{-2 i \omega t} \nonumber\\
&=& -\frac{1}{3} \left[\frac{4 \pi \,\chi_0 N^2}{(2 \pi)^3} \,
{\rm Re}
\int_{|{\bf p}|\leq p_0} d^3p\,\, p_z \,{\rm e}^{i{\bf p}\cdot{\bf x}}\,
 \int \, d\omega \,{\rm e}^{-2 i \omega t} \right] \nonumber\\
&=& -\frac{1}{3}I_{2a}\,. 
\end{eqnarray}
Therefore our result for $I_2$ is 
\begin{equation}
I_2 = I_{2a} + I_{2b} = \frac{2}{3} I_{2a} \,,
\end{equation} 
and so
\begin{equation}
I_2 \approx 
 \frac{\chi_0 N^2}{6 \pi^2} \, g_2(p_0, r) 
\left[\frac{{\rm cos}(2 \Lambda t)- {\rm cos}(2 q p_0 t)}{t}\right] \,,
\end{equation} 
where $g_2(p_0, r)$ is given by Eq.~(\ref{eq:g_2}). 

\begin{figure} 
\begin{center} 
\leavevmode\epsfysize=8cm\epsffile{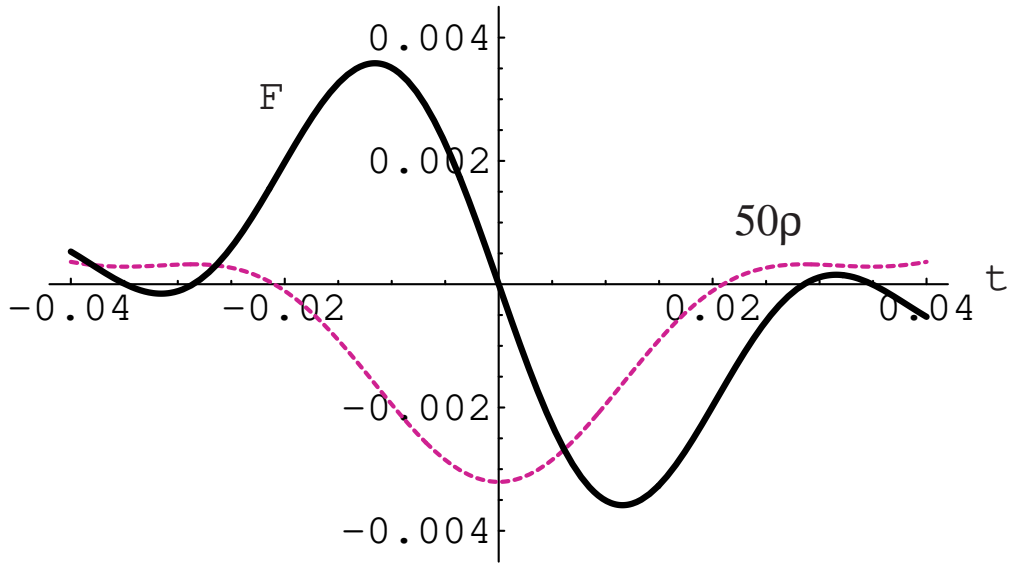} 
\end{center} 
\caption{The energy density, $\rho$, and flux, $F$, at $r=2$ are plotted as 
functions of time. Here we have chosen $p_0=1$, $q=10$, and  
$\Lambda=100$. In order to be visible on this graph, the energy 
density has been multiplied by a factor of $50$. 
Our plot is over a timescale small compared to 
$1/p_0=1$, hence we can ignore the long-timescale dependence of 
$F$ and $\rho$. Note that the sign of the flux changes when the energy density 
reaches its maximum negative value.} 
\label{fig:Flux-ED-Fig1} 
\end{figure}

\begin{figure} 
\begin{center} 
\leavevmode\epsfysize=8cm\epsffile{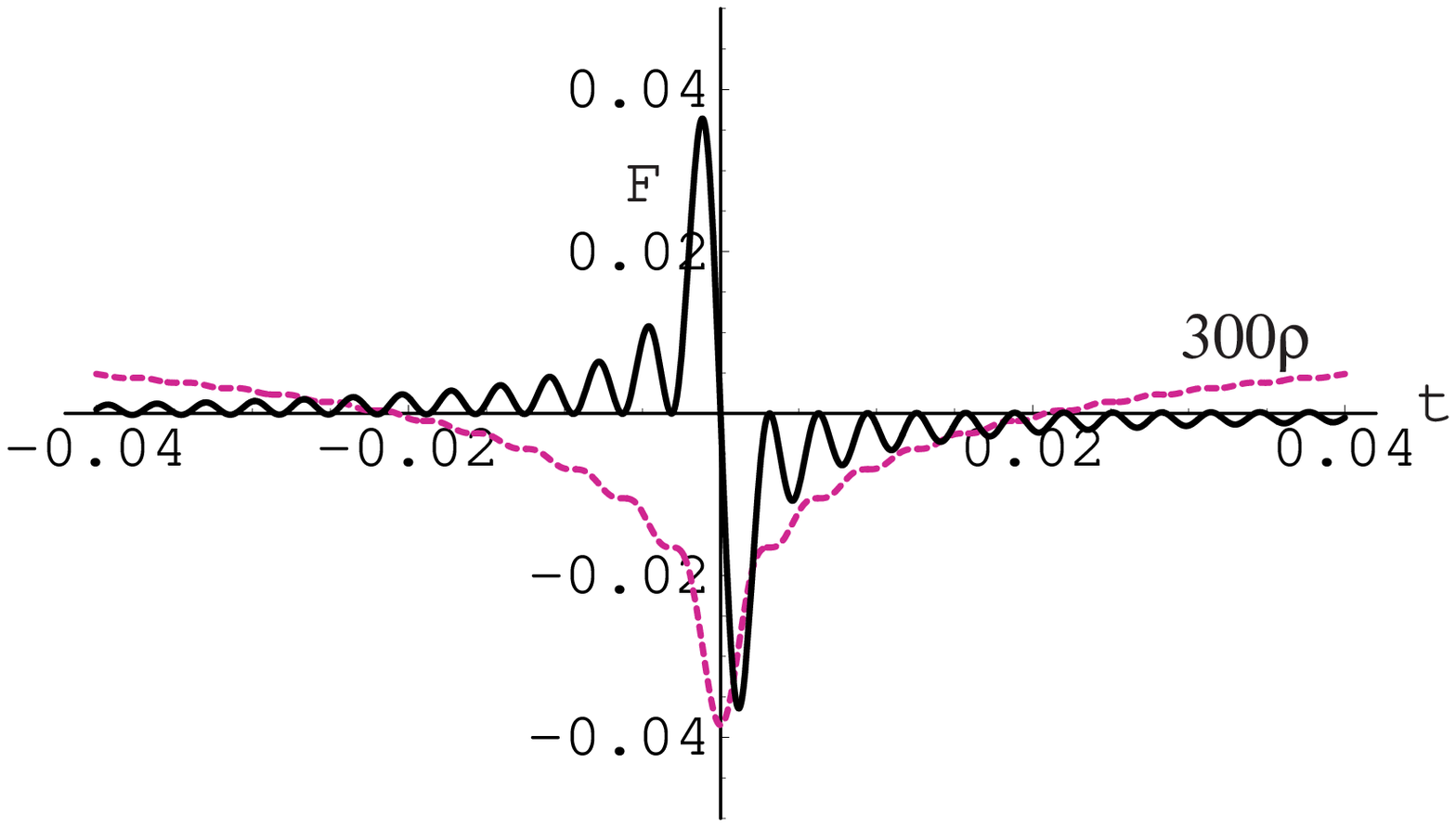} 
\end{center} 
\caption{The energy density and flux at $r=2$ are plotted as 
functions of time, with $p_0=1$, $q=10$, and  
$\Lambda=1000$. In this graph the energy density is multiplied by a factor 
of $300$ in order to be easily visible. Our plot is over the same timescale as 
in Fig.~\ref{fig:Flux-ED-Fig1}. The increase in $\Lambda$ results 
in a higher frequency of oscillation for the flux, $F$. This in turn results 
in the very short timescale fluctuations of the energy density seen in the figure.} 
\label{fig:Flux-ED-Fig3} 
\end{figure}

It is instructive to plot the energy density and flux at constant $r$ 
as a function of time. The flux $F$ is given by 
\begin{equation}
F=\langle T^{0z} \rangle = - \langle T_{0z} \rangle = -I_2 \,.
\end{equation}
In Figs.~\ref{fig:Flux-ED-Fig1} and ~\ref{fig:Flux-ED-Fig3},  
we plot the energy density and flux 
at fixed $r=2$ as a function of time. In both plots, we have 
taken $p_0=1$ and  $q= 10$. Figure~\ref{fig:Flux-ED-Fig1} 
has $\Lambda=100$, while Fig.~\ref{fig:Flux-ED-Fig3} uses
$\Lambda=1000$, resulting in a more rapid variation in time.

In Fig.~\ref{fig:Flux-ED-Fig1}, as we proceed from negative to 
positive values of $t$, we see that the energy density starts out 
approximately constant and positive. It then dips and eventually becomes negative, 
and afterwards becomes positive and constant again at large $t$. Note that 
the dip in the energy density is accompanied by a large positive outward 
flux away from $r=2$, or equivalently a large negative inward flux 
towards $r=2$. The flux goes through zero when the energy density 
reaches its minimum value at $t=0$. It then becomes a large and 
negative outward flux, or equivalently a large and positive inward flux 
towards $r=2$ as the energy density rises and becomes positive again. 
Figure~\ref{fig:Flux-ED-Fig3} shows the more rapid oscillations of the
flux when $\Lambda$ is increased. Note that in this case, the energy
density undergoes small oscillations superposed upon its approach to a
large negative value at $t=0$. 
As originally discussed in 
FHR, the energy density may be made arbitrarily negative over an arbitrarily 
large region at one instant of time. We see here that the accompanying rapidly 
oscillating fluxes will ensure that the quantum inequalities are not violated, 
by quickly ``pumping'' enough positive energy into the region such that the 
time integral of the energy density over any observer's worldline will always be 
positive.

Recall that the spatial region over which the negative energy density
appears can be arbitrarily large. In particular, the light travel time
over this region can be large compared to the time scale on which the 
energy density changes sign. One can see this in our plots by
comparing Fig.~\ref{fig:ED-Fig1} with Fig.~\ref{fig:Flux-ED-Fig3}. The time
scale for the variation of the energy density is about two orders of
magnitude smaller that the size of the negative energy region. Thus
one cannot think of the flow of positive energy into the region as
beginning at the boundary and moving inward in a causal way. Rather,
the energy density and flux are correlated over spacelike
intervals. The energy fluxes depicted in Figs.~\ref{fig:Flux-ED-Fig1}
and \ref{fig:Flux-ED-Fig3} are coherent over the entire region. A
related example of spacelike correlations in quantum states with
negative energy density was discussed in Ref.~\cite{BFR02}.

\section{Energy-Flux Correlations in the Vacuum}

In this section we discuss a different but related issue - the energy density-flux correlations 
in the Minkowski vacuum state. We saw in the previous section that, in the Helfer state, 
the regions of negative energy on a $t=$constant surface were correlated with a rapidly 
time-varying energy flux. Indeed it is this correlation which ensures that the worldline quantum 
inequalities are satisfied. As the energy density in a spatial region became negative, there was 
an accompanying positive energy outflow from (or equivalently, a negative energy inflow into) 
the region. When the energy density in the region began to rise and become positive again, 
this was correlated with a positive energy inflow to (or equivalently, a negative energy 
outflow from) the region. 

Does this behavior represent the general case? One might think that the answer is no. 
The emission of spatially and temporally separated pulses of negative and positive energy 
by moving mirrors would not appear to fit this profile \cite{FD76,FD77,FR-QIC}. However, the 
moving mirror example is two-dimensional. It has been argued elsewhere that, in four-dimensions, 
the emission of two separated plane sheets of negative and positive energy would violate the 
quantum inequalities, unless the positive sheet were infinite in extent \cite{BFR02}. So 
the spatial isolation of negative and positive energy in the moving mirror case may well 
be an artifact of two-dimensions \cite{sphere comment}.

So far all of the scenarios we have discussed involved {\it expectation values} of the energy density and flux. 
In this section, we analyze energy density-flux correlation functions in the Minkowski vacuum state, 
in both two and four dimensions. We find an analogous correlation between energy density and 
flux here as well. That is, if an energy density fluctuation in a region of the vacuum is initially negative, 
it is likely to be followed by a flux fluctuation of positive energy into the region. 
The fact that we see this behavior in the vacuum state as well suggests that it may 
represent the generic case. The correlations already seem 
to be ``coded'' in the vacuum state. 

\subsection{Two Dimensions}

We begin with the calculation of the energy density-flux correlation function for 
a massless scalar field in the Minkowski vacuum state in two dimensions. 
The relevant components of the stress-tensor are
\begin{equation}
T^{xt}=-T_{xt}=-\phi,_x \phi,_t \,,
\end{equation}
and
\begin{equation}
T_{tt}=T^{tt}=\frac{1}{2}[{(\phi,_t)}^2+{(\phi,_x)}^2] \,.
\end{equation}
Renormalized operators, such as $:{(\phi,_t)}^2:$, can be written as 
\begin{equation}
:{(\phi,_t)}^2:={(\phi,_t)}^2-{\langle {(\phi,_t)}^2 \rangle}_0 \, 
\end{equation}
where ${\langle \rangle}_0$ denotes the vacuum expectation value (VEV). 

First use Wick's theorem to write the operator $\phi,_x \phi,_t :{(\phi,_{t'})}^2:$ 
as a sum of normal-ordered products and VEVs: 
\begin{eqnarray}
\phi,_x \phi,_t :{(\phi,_{t'})}^2: &=& :\phi,_x \phi,_t{(\phi,_{t'})}^2:
+ 2:\phi,_t \phi,_{t'}:  {\langle\phi,_x \phi,_{t'} \rangle}_0 \nonumber \\
&+&
2:\phi,_x \phi,_{t'}:  {\langle\phi,_t \phi,_{t'} \rangle}_0 
+2{\langle\phi,_x \phi,_{t'} \rangle}_0 {\langle\phi,_t \phi,_{t'} \rangle}_0 \,,
\end{eqnarray}
where we have also used the fact that $\phi,_x \phi,_t = 
:\phi,_x \phi,_t:$. 
We can get $\phi,_x \phi,_t :{(\phi,_{x'})}^2:$ by letting $t' \rightarrow x'$. 

The flux-energy density correlation function in an arbitrary state is given by
\begin{eqnarray}
C&=&\langle :T^{xt}(x)::T^{t't'}(x'):\rangle =\langle T^{xt}(x):T^{t't'}(x'):\rangle 
\nonumber \\
&=&-\frac{1}{2} \Biggl[\langle\phi,_x \phi,_t :{(\phi,_{t'})}^2:\rangle 
+ \phi,_x \phi,_t :{(\phi,_{x'})}^2: \rangle \Biggr] \nonumber \\
&=&-\frac{1}{2}\Biggl[\langle :\phi,_x \phi,_t {(\phi,_{t'})}^2:\rangle 
+ \langle :\phi,_x \phi,_t {(\phi,_{x'})}^2: \rangle  \nonumber \\
&+& 2\langle:\phi,_t \phi,_{t'}:\rangle {\langle\phi,_x \phi,_{t'}\rangle}_0 
+2\langle:\phi,_t \phi,_{x'}:\rangle {\langle\phi,_x \phi,_{x'}\rangle}_0 \nonumber \\
&+& 2\langle:\phi,_x \phi,_{t'}:\rangle {\langle\phi,_t \phi,_{t'}\rangle}_0 
+2\langle:\phi,_x \phi,_{x'}:\rangle {\langle\phi,_t \phi,_{x'}\rangle}_0 \nonumber \\
&+&  2{\langle\phi,_x \phi,_{t'}\rangle}_0 {\langle\phi,_t \phi,_{t'}\rangle}_0 
+2{\langle\phi,_x \phi,_{x'}\rangle}_0 {\langle\phi,_t \phi,_{x'}\rangle}_0 \Biggr]\,.
\end{eqnarray}
Now consider the case where the given state is the vacuum. Then all of the VEVs 
of normal-ordered products vanish and we have
\begin{equation}
{\langle T^{xt}(x):T^{t't'}(x'):\rangle}_0=
-\left[{\langle\phi,_x \phi,_{t'}\rangle}_0 {\langle\phi,_t \phi,_{t'}\rangle}_0 
+{\langle\phi,_x \phi,_{x'}\rangle}_0 {\langle \phi,_t \phi,_{x'}\rangle}_0 \right]\,,
\end{equation}
with
\begin{equation}
{\langle\phi,_x \phi,_{t'}\rangle}_0= \partial_x \partial_{t'} G = 
\partial_{x'} \partial_{t} G ={\langle\phi,_t \phi,_{x'}\rangle}_0 \,,
\end{equation}
where $G$ is the vacuum two-point function in two-dimensional 
Minkowski space. Therefore our correlation function becomes
\begin{equation}
{\langle T^{xt}(x):T^{t't'}(x'):\rangle}_0=
-{\langle\phi,_x \phi,_{t'}\rangle}_0 
\left( {\langle\phi,_t \phi,_{t'}\rangle}_0 +{\langle\phi,_x \phi,_{x'}\rangle}_0  \right) \,.
\end{equation}

In two-dimensional Minkowski space we have that 
$G=G(\sigma)=  - (4 \pi)^{-1} \,{\rm ln}\, \sigma$, where $\sigma = (1/2){(x^{\mu}-x'^{\mu})}^2 =
(1/2)[{(x-x')}^2 -{(t-t')}^2]$, and $\partial_x G(\sigma) = (x-x') G'(\sigma)$, where 
$G'= \partial G/ \partial \sigma$. Therefore we have that
\begin{eqnarray}
{\langle\phi,_x \phi,_{t'}\rangle}_0&=& \partial_x \partial_{t'} G(\sigma)= (x-x')(t-t')G''(\sigma) 
\nonumber \\
{\langle\phi,_t \phi,_{t'}\rangle}_0&=& \partial_{t'}[-(t-t') G'(\sigma)]
= G'(\sigma)-{(t-t')}^2 \,G''(\sigma) \nonumber \\
{\langle\phi,_x \phi,_{x'}\rangle}_0&=& \partial_{x'}[(x-x') G'(\sigma)]
= -G'(\sigma)-{(x-x')}^2 \,G''(\sigma) \,,
\end{eqnarray}
and so
\begin{equation}
{\langle\phi,_t \phi,_{t'}\rangle}_0+{\langle\phi,_x \phi,_{x'}\rangle}_0 
=-[{(x-x')}^2+{(t-t')}^2]\,G''(\sigma) \,.
\end{equation}
The flux-energy density correlation function in the vacuum then becomes
\begin{equation}
C=\langle T^{xt}(x):T^{t't'}(x'):\rangle = 
(x-x')(t-t')[{(x-x')}^2+{(t-t')}^2] \,{[G''(\sigma)]}^2
\end{equation}

\begin{figure} 
\begin{center} 
\leavevmode\epsfysize=8cm\epsffile{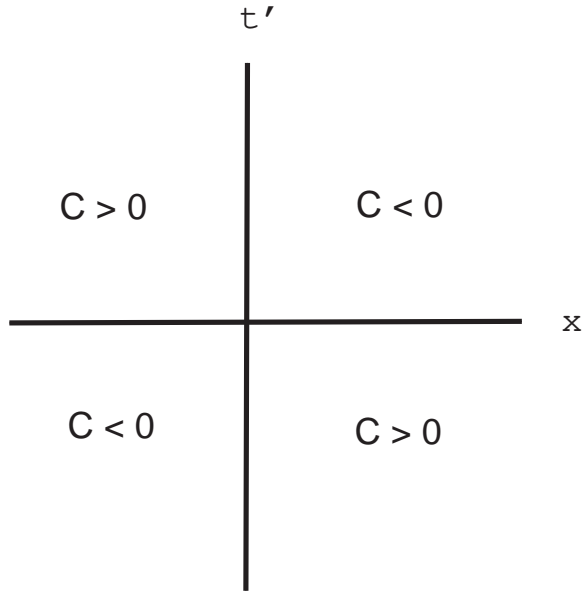} 
\end{center} 
\caption{The sign of $C$ in two dimensions is shown 
in different regions of the $x,t'$ plane.} 
\label{fig:signC2D} 
\end{figure}

Let $t=0$ and $x'=0$, i.e., we look at the correlation between the flux on 
the $t=0$ surface and the energy density on the $x'=0$ worldline. With these choices, 
our correlation function becomes 
\begin{equation}
C=-xt'[x^2+{t'}^2] \,{[G''(\sigma)]}^2 \,.
\end{equation}
The sign of $C$ in different regions of the $x,t'$ plane are shown in 
Fig.~\ref{fig:signC2D}.

\begin{figure} 
\begin{center} 
\leavevmode\epsfysize=12cm\epsffile{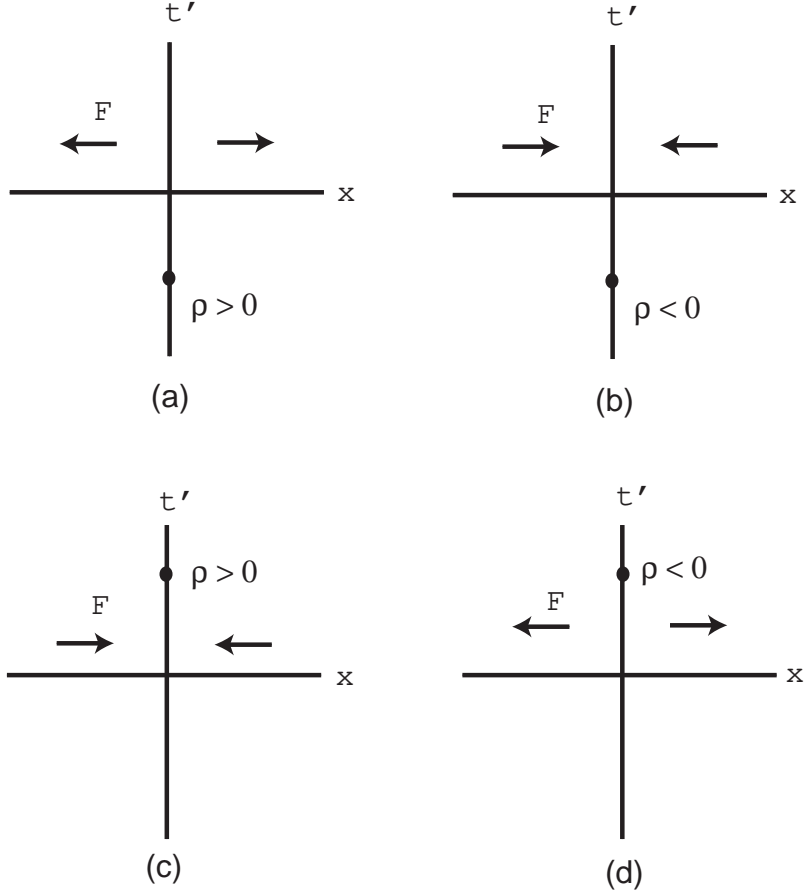} 
\end{center} 
\caption{The correlation of the signs of the energy density and flux
  in two dimensions is shown 
in different regions of the $x,t'$ plane.} 
\label{fig:flux-ed_2D} 
\end{figure}

The correlation of the signs of the energy density, $\rho$, and flux, $F=T^{xt}$, 
is shown in different regions of the $x,t'$ plane in Fig.~\ref{fig:flux-ed_2D}. \hfil\break 

Case A. For $t'<0$ and $x>0$, then $C>0$ and for $t'<0$ and $x<0$, then $C<0$.
Therefore in Fig.~\ref{fig:flux-ed_2D}(a), we see that if $\rho>0$ initially, then $F(x<0)<0$ 
and $F(x>0)>0$ subsequently, on average. In Fig.~\ref{fig:flux-ed_2D}(b), if $\rho<0$ initially, 
then $F(x<0)>0$ and $F(x>0)<0$ subsequently, on average. \hfil\break

Case B.  For $t'>0$ and $x>0$, then $C<0$ and for $t'>0$ and $x<0$, then $C>0$.  
Therefore in Fig.~\ref{fig:flux-ed_2D}(c), we see that if $F(x<0)>0$ 
and $F(x>0)<0$ initially, then $\rho>0$ subsequently, on average. 
In Fig.~\ref{fig:flux-ed_2D}(d), if $F(x<0)<0$ 
and $F(x>0)>0$ initially, then $\rho<0$ subsequently, on average. \hfil\break
Thus in all cases, the sign of the energy density at $x'=0$ and the fluxes at $t=0$ 
are correlated as we would expect.

\subsection{Four Dimensions}
\begin{figure} 
\begin{center} 
\leavevmode\epsfysize=5cm\epsffile{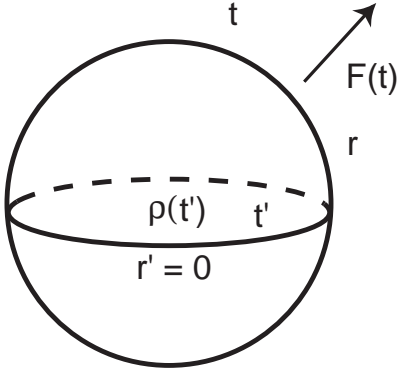} 
\end{center} 
\caption{Here the relation between the energy density at time $t'$ 
and $r'=0$ and the flux $F$ at time $t$ and radius $r$ is sketched.
A positive energy density at an earlier time $t'$ is correlated with
an outgoing flux at a later time $t$, and {\it vice versa}.} 
\label{fig:flux-ed_4D} 
\end{figure}

We now compute the vacuum energy density-flux correlation function in four-dimensional 
Minkowski space. The radial flux is given by 
\begin{equation}
F=T^{rt}=\phi^{,r}\phi^{,t}=-T_{rt}=-\phi,_r \phi,_t \,,
\end{equation}
and the energy density by 
\begin{equation}
\rho=T^{tt}=\frac{1}{2}(\phi,_t \phi,_t + \phi,_i \phi^{,i}) \,.
\end{equation}
For the correlation function we have
\begin{equation}
C=\langle :T^{rt}(x)::T^{t't'}(x'):\rangle =\langle T^{rt}(x):T^{t't'}(x'):\rangle \,.
\end{equation}
A calculation analogous to the one given for two dimensions yields
\begin{eqnarray}
C&=&-\frac{1}{2} {\langle :\phi,_r \phi,_t: :(\phi,_t \phi,_t + \phi,_i \phi^{,i}):\rangle}_0
\nonumber \\
&=& -{\langle \phi,_r \phi,_{t'} \rangle}_0 {\langle \phi,_t \phi,_{t'} \rangle}_0 - 
{\langle \phi,_r \phi,_{i'} \rangle}_0 {\langle \phi,_t \phi^{,i'})\rangle}_0 \,.
\end{eqnarray}
Here the vacuum two-point function is $\langle \phi(x) \phi(x') \rangle = G(\sigma)=1/(8 \pi^2 \sigma)$, 
where $\sigma = (1/2) [-{\Delta t}^2 + {|\Delta {\bf x}|}^2]$, and 
$\Delta t = t-t'$, $\Delta {\bf x}={\bf x}-{\bf x'}$. So we have that
\begin{equation}
{\langle \phi,_t \phi,_{t'} \rangle}_0 =\partial_t \partial_{t'} G(\sigma) 
=\sigma,_{tt'}G'(\sigma)+G''(\sigma) \sigma,_t \sigma,_{t'}\, = 
G'(\sigma)-{(\Delta t)}^2 \, G''(\sigma) \,,
\end{equation}
and
\begin{equation}
{\langle \phi,_r \phi,_{t'} \rangle}_0 =\partial_r \partial_{t'} G(\sigma) 
=\sigma,_{rt'}G'(\sigma)+ G''(\sigma) \sigma,_r \sigma,_{t'}\, = 
\Delta t\,\sigma,_r\, G''(\sigma) \,.
\end{equation}
Furthermore,
\begin{eqnarray}
{\langle \phi,_t \phi^{,i'} \rangle}_0 &=& -\Delta t \,\sigma^{,i'} \,G''(\sigma) \,,
\nonumber \\
{\langle \phi,_r \phi,_{i'} \rangle}_0 &=& \sigma,_{i'r} \,G'(\sigma)+
\sigma,_{i'} \sigma,_{r}\,G''(\sigma) \,,
\end{eqnarray}
so
\begin{equation}
{\langle \phi,_r \phi,_{i'} \rangle}_0 {\langle \phi,_t \phi^{,i'} \rangle}_0
=-\Delta t\,G''(\sigma) \, \left[(\sigma^{,i'} \, \sigma,_{i'r})\,G'(\sigma)
+ (\sigma^{,i'} \, \sigma,_{i'}) \,\sigma,_{r} \,G''(\sigma) \,.  \right]
\end{equation} 
We need to calculate the derivatives of $\sigma$ which appear in the
above expression. The scalar quantity $\sigma^{,i'} \, \sigma,_{i'}$
is most easily computed in Cartesian coordinates, where $\sigma,_{i'}=
-\Delta x^i$ and hence  $\sigma^{,i'} \, \sigma,_{i'} = 
|\Delta {\bf  x}|^2$. We consider the case when ${\bf x'}=0$. We then find 
\begin{eqnarray}
\sigma^{,i'}\sigma,_{i' r} &=& r \nonumber \\
\sigma,_r &=& r \nonumber \\
\sigma^{,i'}\sigma,_{i'} &=& r^2 \,.
\end{eqnarray}
As a result our energy density-flux vacuum correlation function reduces to
\begin{equation}
C=r \,{\Delta t} \left[ {({\Delta t} )}^2 + r^2 \right] \, {[G''(\sigma)]}^2 \,,
\end{equation}
where $\sigma=(1/2)[r^2-{({\Delta t} )}^2]$ for $r'=0$. Note that the sign of $C$ 
depends only on the sign of $\Delta t$. Recall that ${\Delta t} = t-t'$. 
If $t>t'$, then $C>0$. Consequently, $\rho(t')>0$ implies
$F=T^{rt}(t)>0$ on average,
as illustrated in Fig.~\ref{fig:flux-ed_4D}, and   
$\rho(t')<0$ implies $F=T^{rt}(t)<0$ on average. This is consistent with our two-dimensional results. 
If $\rho >0$ at the origin at one time, the flux at a later time is likely to be outgoing. Similarly, 
if $\rho<0$ at one time, the later flux is more likely to be ingoing. 
We can also reverse the time order. An outgoing flux at $t$ is correlated with $\rho<0$ at $t'>t$, 
and an ingoing flux at $t$ is correlated with $\rho>0$ at $t'>t$.

Lastly we discuss the falloff of the correlations in both two and four dimensions. 
In two dimensions, choosing $t=0$, we have ${\Delta t}=t-t'=-t'$ and so 
\begin{equation}
C=-x t' \, (x^2+{t'}^2) \,{[G''(\sigma)]}^2  \,.
\end{equation}
Since in two dimensions $G=-(4 \pi)^{-1}\,{\rm ln}\, \sigma$, $G''=
1/(4\pi \sigma^2) \propto -1/{(x^2-{t'}^2)}^2$. 
If we let $x \rightarrow \infty$, with $t'$ fixed, then 
\begin{equation}
C \sim -\frac{t'}{x^5} \,.
\end{equation}
Alternatively, if we let $t' \rightarrow \infty$, with $x$ fixed, then 
\begin{equation}
C \sim -\frac{x}{{(t')}^5} \,.
\end{equation}
Thus the falloff is symmetrical for both spacelike and timelike separations. 

In four dimensions, with $r'=0$ and since $G = 1/(8 \pi^2\sigma)$, 
$G'' = 1/(4 \pi^2 \sigma^3) \propto 1/{[ r^2-{({\Delta t})}^2 ]}^3$, we have that 
for $r \rightarrow \infty$, with $\Delta t$ fixed, then 
\begin{equation}
C \sim \frac{\Delta t}{r^9} \,.
\end{equation}
For ${\Delta t} \rightarrow \infty$, with $r$ fixed,  
\begin{equation}
C \sim \frac{r}{{(\Delta t)}^9} \,.
\end{equation}
and again the falloff is symmetric between space and time. Note also 
that $C$ has the same form in two and four dimensions, apart from the 
difference in the expression for $G(\sigma)$.

\section{Conclusions} 
In this paper we have examined in detail the behavior and correlation 
of energy density and flux in a ``Helfer state''. This state, 
first suggested by Adam Helfer, is one in which the energy density 
can be made arbitrarily large and negative over an arbitrarily large region of 
space, even in Minkowski spacetime. However, this can be achieved only at one 
instant in time. For the quantum inequalities to hold, there must be large 
time-varying fluxes associated with these states. We show that this is in fact the case, 
and analyze the correlated behavior of the expectation values of the 
flux and energy density, in order to gain 
a deeper insight into these remarkable quantum states. 

In the second part of the paper, we examine the energy density-flux correlation 
function in the Minkowski vacuum state. Although we now work with correlation functions, 
as opposed to expectation values, we find qualitatively similar behavior for energy 
density and flux fluctuations. An initial negative energy 
vacuum fluctuation in some region of space is correlated with a subsequent flux fluctuation 
of positive energy into the region. 

The quantum inequalities place a (negative) lower bound 
on the distribution function for energy density vacuum fluctuations \cite{Chris_comment}. 
Perhaps the mechanism which ensures that the quantum inequalities hold in the Helfer state, 
as well as in other quantum states associated with negative energy, is, at least in some 
sense, already ``encoded'' in the fluctuations of the vacuum. 

\begin{acknowledgments}
This work was supported in part by the National
Science Foundation under Grants PHY-0555754 and PHY-0652904.
\end{acknowledgments}


\begin{thebibliography}{28}
\bibitem{sphere comment} A more appropriate analog in four-dimensions might be 
two concentric shells of positive and negative energy as discussed in 
Ref.~\cite{BFR02}.
\end{thebibliography}
\end{document}